\documentstyle[12pt]{article}
\begin{document}
\title{\bf Commensurate-Incommensurate Phase Transitions for Multi-Chain
Quantum Spin Models: Exact Results}
\vspace{1.5em}
\author{A. A. Zvyagin \\
B.~I.~Verkin Institute for Low Temperature Physics and Engineering \\
of the National Academy of Sciences of Ukraine, \\
47 Lenin Avenue, Kharkov, 61164, Ukraine}

\date{August, 18, 1999}
\maketitle
\begin{abstract}
The behavior in an external magnetic field is studied exactly for a wide 
class of multi-chain quantum spin models. It is shown that the magnetic field
together with the inter-chain couplings cause the commensurate-incommensurate
phase transitions between the gapless phases in the groundstate. The 
conformal limit of these models is studied and it is shown that the low-lying
excitations for the incommensurate phases are not independent, because they
are governed by the same magnetic field (chemical potential for excitations). 
A scenario for the transition from one to two space dimensions for the 
exactly integrable multi-chain quantum spin models is proposed and it is 
shown that the incommensured phases in an external magnetic field disappear 
in the limit of an infinite number of the coupled spin chains. The 
similarities in the external field behavior for the quantum multi-chain spin 
models and a wide class of quantum field theories are discussed. The scaling 
exponents for the appearence of the gap in the spectrum of the low-lying 
excitations of the quantum multi-chain models due to the relevant 
perturbations of the integrable theories are calculated.
\end{abstract}

\vspace{1.5em}

\centerline{PACS numbers: 75.10.Jm,11.10.Kk}
\newpage

\section{\bf Introduction}

There has recently been considerable interest on low-dimensional quantum
correlated spin and electron systems. These systems, especially
one-dimensional (1D), manifest the specific features of, e.g., magnetic
behavior at low temperatures, which are absent for the standard, conventional
3D magnetic systems. Spin systems usually manifest 1D behavior for the
temperatures higher than the temperature of the 3D magnetic ordering, but
lower than the maximum characteristic energy of the interaction between spins,
i.e. in our case the intra-chain spin-spin coupling. The origin of such
specific features is the enhancement of the quantum fluctuations of the 1D
systems due to the peculiarities of the 1D density of states together with
the quantum nature of spins.

Moreover, during the last decade a large number of new quasi 1D spin
compounds were created and studied experimentally. These compounds manifest
at low temperatures the properties of a single or several quantum spin chains
weakly coupled to each other \cite{ER,exp}. It is strongly believed that this
class of compounds will provide the new information on the transition from 1D
to 2D in quantum many-body physics. It is very important, because the 2D
quantum many-body physics has been a challenge for both theorists and
experimentalists since the beginning of the study of low dimensional quantum
systems. On the other hand, the advantage of the 1D theoretical studies is the
possibility of obtaining exact solutions by using non-perturbative methods,
which are difficult to apply for the higher-dimensional quantum many-body
models. The results of the exact calculations of the 1D models can serve as
testing grounds for the use of perturbative and numerical methods in more
realistic situations.

Recently several exactly solvable models \cite{Tsv,PopZv,MT} have been 
introduced, in which the zigzag-like interaction between {\em two} quantum 
spin chains was studied exactly using the Bethe {\em ansatz} technique 
\cite{Bet}. This method is widely known by now, see e.g., the recent 
monography \cite{qism} and references therein. The Bethe ansatz method 
permits exact calculation of the static characteristics of quantum many-body 
systems, such as the groundstate behavior, the influence of an external 
magnetic field, and the thermodynamic features of e.g., the temperature 
dependencies of the specific heat, magnetic susceptibility, etc. These 
results should apply to more realistic systems, but it is not obvious how the
interactions between the chains modify the answers. The mean-field like
approximations for the inter-chain couplings are not sufficient, because the
mean field approach in any version already implies the existence of the
(sometimes hidden) order parameter. It is, unfortunately, also unclear 
whether the numerical calculations, which can be directly applied for the
quantum many-body systems of very small sizes by now (say, at most several
tens of sites) describe well the properties of the real systems, in which,
even in quasi-1D ones, the number of sites is at least of order of 10$^{8}$
or higher. On the other hand, it must be admitted that some features of the
exactly solvable 1D models are far from what is observed experimentally, but 
these unrealistic features of the 1D models are known and simple to
recognize.

The behavior of the multi-chain spin systems in an external magnetic field is
especially interesting, see e.g., \cite{OYA,Zv3,FR,MT} because of (i) the
possibility of experimental observations due to recent progress in the
high magnetic field measurements, and (ii) very interesting theoreticallly
predictable effects which are possible to recognize in experiments, such as 
phase transitions in the external magnetic field. However several important
issues are far from being solved in the quantum two-chain spin models. For
example, there are three questions that need to be answered: (1) Are the
properties of those exactly solvable two-chain spin models unique or it is
possible to say something about the more general class of two-chain quantum
spin models? (2) How are the multi-chain quantum models connected to the 2D
many-body systems, i.e. what is the scenario of the transition from 1D to 2D
when one increases the number of coupled chains while keeping the conditions of
integrability? and (3) What will happen with the behavior of the
non-integrable multi-chain spin models if one goes beyond the framwork of 
integrability i.e. adding some perturbations to the exactly solvable model? 
(For example, Ref.~\cite{FR} implies that namely the spin chirality, which
separatly breaks the time-reversal and parity symmetries in the two-chain
integrable model \cite{Zv1}, is the reason for the emergence of the
additional phase transitions in an external magnetic field for the two-chain
spin ${1\over2}$ model as compared to the single-chain system).

The goal of this paper is to answer these questions. First, we re-visit the
exactly integrable two-chain spin ${1\over2}$ model and show that the
inclucion of the {\em magnetic anisotropy} of the ``easy-plane'' type, with
which the system stays in the quantum critical region, will not drastically 
change the behavior in an external magnetic field but {\em will shift the
critical values} of the magnetic fields and intra-chain couplings at which
the phase transitions occur and will affect the critical exponents. We will
show that these two-chain spin models {\em share the most important features}
of the behavior in an external field with the wide class of the (1+1) {\em
quantum field theories}. Next, we will {\em introduce the higher-spin}
realizations of the two-chain spin models, e.g., investigating the important
class of 1D two-chain {\em quantum ferrimagnets} with different spin values
at the sites of each chain. We will also investigate the behavior of the
{\em multi-chain} exactly solvable spin models in an external magnetic field 
and show how the additional phase transitions arising due to the
increasing number of chains vanish in the quasi-2D limit. Finally, we will
show {\em how} the relevant deviations from the integrability, e.g., the
absence of the terms in the Hamiltonian which separately break the parity and
time-reversal symmetries give rise to {\em gaps} in the spectra of low-lying 
excitations of the multi-chain quantum spin systems and we will calculate the 
scaling exponents for the gaps.

The paper is organized as follows. In Section~2 we re-visit the exactly
solvable two-chain uniaxial spin model \cite{PopZv} to remind the reader of 
the main steps of the Bethe ansatz. The investigations of
Refs.~\cite{Zv3,FR} of isotropic spin ${1\over2}$ two-chain models are
generalized in this section for the case of uniaxial magnetic anisotropy. The
calculations in this section are rather simple, but we write them in detail
because they provide the basis for the more nontrivial generalizations of
this class of models, and will be used in the following Sections. In
Section~3 we point out the similarities between the behavior of the
uniaxial two-chain quantum spin models and a class of quantum field theories
(QFT) in an external magnetic field, predicting new phases for the QFT. In
Section~4 we introduce the $SU(2)$ generalization of the integrable
two-chain model for higher values of the site spins (possibly different) in
each chain, i.e., quantum ferrimagnet. We point out the similarities of the
quantum ferrimagnet with the QFT with a nonzero Wess-Zumino term and
predict new phases for the latter in an external magnetic field. We
derive the integral eguations for the critical exponents. In Section~5 we
consider the multi-chain quantum spin model and discuss how the external
field behavior of the integrable multi-chain models is changed when the 
number of chains is increased while preserving the exact solvability. In
Section~6 we briefly sketch how the deviations from integrability change the
magnetic and low-temperature properties of this class of multi-chain quantum
spin systems. The paper is closed with a discussion of the main results
and some conclusions.

\section{\bf Two-chain uniaxial quantum spin model}

A common property of some of the Bethe ansatz solutions is the presence of 
shifts $\theta_j$ of the spectral parameter $\lambda$ for the associated 
transfer matrix of an algebraic version of the Bethe ansatz (the Quantum 
Inverse Scattering Method, [QISM] \cite{qism}). Those shifts also appear in 
the Bethe ansatz equations (BAE) for the quantum numbers called {\em 
rapidities}, which parametrize the eigenfunctions and eigenvalues of the 
Hamiltonians. Hence, the distributions of the rapidities are also affected by 
the shifts. An interesting property is connected with those shifts: depending 
on their values and the external magnetic field, even for (quasi)particles of 
the same type, additional minima may appear in distributions of the 
rapidities. These additional minima also result in the nonmonotonic behavior 
of the dispersion laws of the low-lying excitations. Also, they provide 
additional Dirac seas for low-lying excitations, changing the structures of 
the physical groundstates of the models. These additional minima determine 
the special behavior of the models in the external magnetic field 
\cite{Tsv,Zv3,FR,MT}. In particular, the appearence of the new phases and new 
phase transitions is namely due to the emergence of these new minima in the 
distributions of the quantum numbers.

To set the stage, let us first remind the reader about the main steps of the 
QISM. The common feature of the Bethe ansatz solvable models is the
factorization of a monodromy matrix (the ordered product of all two-particle
scattering matrices, which depend on some spectral parameter) \cite{qism}.
Exact (Bethe ansatz) integrability requires exclusively {\em elastic}
scattering between (quasi)particles. For such a theories two-particle
scattering matrices and $L$-operators satisfy the Yang-Baxter relation
\cite{Bax,qism}. In turn, the factorization of the monodromy matrices garantees
that they satisfy the Yang-Baxter equations, too. The transfer matrices of the
associated statistical problem are traces over some additional, auxiliary 
subspace, of monodromy matrices \cite{qism}. The most important feature of
transfer matrices with different spectral parameters is their commutativity.
The necessary and sufficient condition for this is the validity of the
Yang-Baxter equations for two-particle scattering matrices and hence for
monodromy matrices. The commutativity of transfer matrices implies that one
can construct an infinite number of integrals of motion, which commute with 
one another and with the transfer matrix. Therefore the exact integrability is
proved. Usually the structure of these integrals of motion is determined by
their {\em locality}. For instance, the best known of series of integrals
of motion is the series of derivatives with respect to the spectral parameter
of the logarithm of a transfer matrix taken at some special value of former
\cite{qism}. Locality means that for the first derivative of the logarithm of
the transfer matrix (usually called the Hamiltonian of the lattice system)
only short-range particle-paricle interactions contribute.

In this paper we will see that namely the aforementioned shifts of the spectral
parameters yield new phases in the groundstate behavior in an external
magnetic field of a wide class of exactly solvable models, quantum spin
multi-chain models and QFT. We will show that in the conformal limit these
phases of the {\em lattice models} correspond to {\em one} Wess-Zumino-Witten
(WZW) model or to {\em several} of them with {\em dressed charges} 
(proportional to the compactification radii) of scalar or matrix types for 
each of the phases, respectively.

Let us start with the form of the Bethe ansatz equations (BAE) for the set of
rapidities $\{ u_{\alpha}\}_{\alpha =1}^M$. In this paper we will concentrate
only on the critical, ``easy-plane'' type of the magnetic anisotropy for the
antiferromagnetic spin multi-chain models, $0 \le \gamma \le \pi/2$ ($\gamma
= \pi /q$, $q$ integer, parametrizing the magnetic anisotropy), and the
{\em repulsive} interactions in QFT. This correspond to hyperbolic or
rational solutions of the Yang-Baxter equations for two-particle scattering
matrices, or to $U(1)$ and $SU(2)$ symmetries of the scattering processes,
respectively. For the simplest case of one shift $\theta$, which is connected
to the two-chain quantum spin models and most of QFT, the BAE have the form
(here we use more general hyperbolic parametrization first; for the rational
limit see below) \cite{PopZv}:
\begin{equation}
\prod_{\pm}e_1^{N_{\pm}}(u_{\alpha} \pm \theta) = e^{i\pi M}
\prod_{\beta=1,\beta \ne \alpha}^M e_2 (u_{\alpha} - u_{\beta}) \ ,
\label{BAE}
\end{equation}
where $N_{\pm}$ are the numbers of sites in each of spin chains, $e_n (x) =
\sinh (x + i\gamma {\textstyle \frac {n}{2}})/\sinh (x - i\gamma {\textstyle
\frac {n}{2}})$ and $M$ is the number of down spins. The shift $\theta$
determines the inter-chain coupling constant for two-chain quantum spin
${1\over2}$ models \cite{PopZv,Zv1,Zv2}. Please note that the Bethe ansatz
equations are just the {\em quantization conditions} for the rapidities,
which parametrize the eigenwaves and eigenvalues of the many-body quantum
model. The Hamiltonian is the first derivative of the logarithm of the
transfer matrix (note that the transfer matrix of the two coupled spin chains
in this integrable model is the product of two ``standard'' transfer matrices
of each chain with the spectral parameters $\lambda \pm \theta$ \cite{Zv1}):
\begin{eqnarray}
&&{\hat H}_{1/2} = {1\over \sinh^2 \theta + \sin^2 \gamma }\sum_n \biggl( 
\sinh^2 \theta {\hat E}\bigl({\vec S}_{n,1}{\vec S}_{n+1,1} +
{\vec S}_{n,2}{\vec S}_{n+1,2} \bigr) +
\nonumber \\
&&2\sin^2 \gamma {\hat I}
{\vec S}_{n,1}\bigl( {\vec S}_{n,2} + {\vec S}_{n+1,2}\bigr) +
2\sin \gamma \sinh \theta \bigl(
{\hat J}{\vec S}_{n+1,2}-{\hat J}{\vec S}_{n,1}\bigr)
\bigl[{\vec S}_{n+1,1}\times {\vec S}_{n,2} \bigr] \biggr) \ ,
\label{biHam}
\end{eqnarray}
where $diag (a,b,c)$ is $3\times3$ diagonal matrix, 
\begin{eqnarray}
&&{\hat E} = diag \bigl(1, 1, \cos \gamma \bigr) \ , 
\nonumber \\
&&{\hat I}= diag \bigl( \cosh \theta, \cosh \theta, \cos \gamma \bigr) \ , 
\nonumber \\
&&{\hat J} = diag \bigl( \cos \gamma, \cos \gamma, \cosh \theta \bigr) \ , 
\nonumber 
\end{eqnarray}
and $[.\times .]$ denotes the vector product. Please note 
that the sum runs over $n$ to $N_+$ for the chain with spins $S_{n,1}$ and to 
$N_-$ for the chain with spins $S_{n,2}$. The parameter $\theta$ determines 
the intra-chain coupling in our two-chain spin model. For $\theta=0$ the 
Hamiltonian and BAE coincide with the ones for the single ``easy-plane'' 
antiferromagnetic spin ${1\over2}$ chain of length $N_+ +N_-$ with the only 
nearest neighbour interactions in it. The eigenvalue of the Hamiltonian 
(energy) is parametrized as the function of the rapidities as follows:
\begin{equation}
E=\sin \gamma \sum_{\pm}\sum_{\alpha=1}^{M}N_{\pm}(e_1(u_{\alpha} \pm \theta)
+ e_1^{-1}(u_{\alpha} \pm \theta)) + E_0 \ ,
\label{energy}
\end{equation}
where $E_0$ is the energy of the vacuum (ferromagnetic) state (with $M=0$).
The isotropic $SU(2)$-symmetric antiferromagnetic quantum spin two-chain model
\cite{Zv1,Zv2,Zv3,FR} can be obtained from the uniaxial ($U(1)$-symmetric)
one of Eqs.~(\ref{BAE})-(\ref{energy}) by the simple change of variables in
the limit: $u_{\alpha} \to \gamma u_{\alpha}$, $\lambda \to \gamma \lambda$,
$\theta \to \gamma \theta$, $\gamma \to 0$. (The last limit corresponds to
the rational, $SU(2)$-symmetric solution of the Yang-Baxter equations for
two-particle scattering matrices). The two-chain isotropic
($SU(2)$-symmetric) spin ${1\over2}$ Hamiltonian obtained in this limit from
Eq.~(\ref{biHam}) takes the form \cite{PopZv,Zv1,Zv2,Zv3,FR}:
\begin{eqnarray}
&&{\hat H}_{is} = {1\over 1+\theta^2}\sum_n \biggl( \theta^2
\bigl({\vec S}_{n,1}{\vec S}_{n+1,1} + {\vec S}_{n,2}{\vec S}_{n+1,2} \bigr)
+ 2{\vec S}_{n,1} \bigl( {\vec S}_{n,2} + {\vec S}_{n+1,2} \bigr) +
\nonumber \\
&&2\theta \bigl( {\vec S}_{n+1,2} - {\vec S}_{n,1}\bigr) \bigl[
{\vec S}_{n+1,1}\times {\vec S}_{n,2}\bigr] \biggr) \ . \
\label{isHam}
\end{eqnarray}
The summations over $n$ runs to $N_{\pm}$ for each kind of spins, 
respectively. Note that for $\theta \to \infty$ Eqs.~(\ref{isHam}) and BAE 
recover the Hamiltonian and BAE of two {\em decoupled} spin ${1\over2}$ 
chains of lengths $N_{\pm}$ with the only nearest neighbour interactions in 
each of the chains.

The solution to the BAE Eqs.~(\ref{BAE}) is usually obtained in the 
thermodynamic limit ($N_{\pm},M \to \infty$, with the ratio $M/(N_++N_-)$ 
fixed). Here instead of the discret set of rapidities one introduces the 
distribution of a continuous density of rapidities. The groundstate 
corresponds to the solutions of the BAE with negative energies, i.e., it is 
connected with the filling up the Dirac sea(s) for the model. For the 
``easy-plane'' antiferromagnetic two-chain spin ${1\over2}$ model the 
groundstate corresponds to the filling of the Dirac sea for the {\em real} 
rapidities, i.e., no spin boundstates have negative energies. In the 
thermodynamic limit the real roots of Eqs.~(\ref{BAE}) are distributed 
continuously over some intervals, which determine the Dirac seas of the 
model. The set of integral equations for the dressed densities of rapidities 
$u_{\alpha}$ ($\rho (u)$) and dressed energies of low-lying quasiparticles 
($\varepsilon (u)$) are (see, e.g., Ref.~\cite{qism} for the standard 
procedure of deriving these integral equations from the BAE and 
Refs.~\cite{Zv1,Zv2} for the isotropic two-chain spin ${1\over2}$ model):
\begin{equation}
\rho (u) + \int_{(Q)} dv K(u-v) \rho (v) = \sum_{\pm}{\frac {N_{\pm}}{N}}
\rho_{\pm}^0
\label{rho}
\end{equation}
and
\begin{equation}
\varepsilon (u) + \int_{(Q)} dv K(u-v) \varepsilon (v) = h -
\sum_{\pm}{\frac {N_{\pm}}{N}}\varepsilon_{\pm}^0 \ , \
\label{eps}
\end{equation}
where the kernels of integral equations are
\begin{equation}
K(u)= {\partial \ln e_2(u)\over \partial u} = { \sin (2\gamma) \over 2\pi
[\cosh (u) - \cos (2\gamma)]}\ . \
\label{kernel}
\end{equation}
and $h$ is an external magnetic field. The values
\begin{equation}
\rho_{\pm}^0(u) = {\partial \ln e_1(u \pm \theta )\over \partial u}
\equiv {\partial p_{\pm}^0(u) \over \partial u} =
{\sin \gamma \over 2\pi [\cosh (u\pm \theta) - \cos (\gamma)]}
\label{rho0}
\end{equation}
are bare densities of the rapidities, and
\begin{equation}
\varepsilon_{\pm}^0(u) = h - {\sin^2 \gamma \over \cosh (u \pm \theta)
- \cos (\gamma)}
\label{eps0}
\end{equation}
are bare energies (here ``bare'' corresponds to noninteracting particles, and
the interaction ``dresses'' them as usual \cite{qism}). The integrations are
performed over the domain $(Q)$, determined in such a way that the dressed
energies inside these intervals are negative. The limits of integrations are
determined by the zeros of the dressed energies, and are the Fermi points for
each sea. The analysis of the integral equations Eqs.~(\ref{rho}),(\ref{eps})
in an external magnetic field shows that in general, for some values of
$\theta$ and $h$, there can be one Dirac sea (it corresponds to one minimum
of the bare density of rapidities and, hence to one minimum of the bare 
energy). On the other hand, for higher values of $\theta$ and for some domain 
of $h$ two Dirac seas of the same type of (gapless, see below) excitations 
are possible (for two minima of the bare energies of the rapidities and thus 
two minima of the bare density). Note that for $\theta \to \infty$ at fixed 
$N_{\pm}$ all the roots of the integral BAE separate into two sets of 
``right-'' and ``left-moving'' seas, centered at $\pm \theta$, respectively.

Here we briefly re-visit the analysis of Refs.~\cite{Zv3,FR}, but for
the case of the {\em uniaxial} two-chain model. Analytic solutions to
Eqs.~(\ref{rho})-(\ref{eps}) can be easily obtained in closed form in the
limit of zero field and equal lengths of the chains $N_+ = N_-$. The simplest
non-trivial exited quasiparticle (spinon) is a hole in the Dirac sea for real
rapidities with the quasimomentum
\begin{equation}
p(u_0) = 2\arctan \biggl( {{\sinh (\pi u_0 / \gamma)}\over \cosh
(\pi \theta /\gamma)} \biggr) \ , \
\label{Psp}
\end{equation}
where $u_0$ is the spinon's rapidity. Note that due to topological reasons
such particles have to exist in pairs for the $SU(2)$-symmetric case, etc.
\cite{AL1,FT}. The energy of this spinon is given by
\begin{equation}
\epsilon (u_0) = -\sin \gamma {\partial p(u_0)\over \partial u_0} \ . \
\label{Esp}
\end{equation}
It can be rewritten as function of the quasimomentum, i.e., in the form of the
commonly used dispersion law
\begin{equation}
\epsilon (p) = {\pi \over \gamma }\sin \gamma \tanh {\pi \theta \over \gamma }
\sin {p \over 2}\biggl[ \cos^2 {p \over 2} + \sinh^{-2}{\pi \theta \over
\gamma }\biggr]^{1/2} \ . \
\label{disp}
\end{equation}
A spinon corresponds in the usual Bethe ansatz classification of BAE solutions
to a string of length 1 \cite{qism}. Naturally Eqs.~(\ref{BAE}) have string
solutions of higher lengths too. Other spin excitations can be obtained as
combinations of spinon quasiparticles and higher-length strings with
different rapidities. However, spinons here are picked out because only their
dressed energies may be negative, i.e., only spinons may form Dirac seas of
the groundstate of the model.

One can see that the dispersion law Eq.~(\ref{disp}) of the low-lying
excitation of the ``easy-plane'' two-chain spin ${1\over2}$ antiferromagnetic
model is factorized into two parts: the gapless part at $p = 0,\pi$ and the
gapful one at $p = \pi/2$, cf. \cite{Zv3,FR}. The former corresponds to the
oscillations of the magnetization, while the latter is connected with the
oscillations of the staggered magnetization \cite{Zv3}. The analysis, similar
to the one of the solutions of Eqs.~(\ref{rho}),(\ref{eps}) for nonzero
magnetic field $h \ne 0$ (here we point out that according to the very
accurate analysis \cite{FKM} the solution of the integral BAE in the first
order approximation reproduces correctly both low- and high-coupling
asymptotic behavior) shows that: (i) the dressed energy of a spinon as a 
function of the dressed quasimomentum has only one extremum, a maximum at $p 
= \pi /2$ for $\theta < \theta_c$ and (ii) for $\theta > \theta_c$ there are 
two maxima and one minimum (situated at $p = \pi /2$). At the (tri)critical 
point $\theta_c$, the minimum disappears and two maxima joint into one more 
flat (at $p = \pi /2$). In the limit $\theta \to \infty$ the mimimum is 
transformed into a cusp. It reveals that the gap of the staggered 
magnetization vanishes in this limit of two independent spin chains. This 
simple picture helps us to understand what happens if one switches on an 
external magnetic field $h$. Besides the usual phase transition to the 
ferromagnetic (spin-polarized) phase at
\begin{equation}
h_s = \sum_{\pm} {\frac {N_{\pm}}{N}}\varepsilon_{\pm}^0(0)
\label{hs}
\end{equation}
there is an additional transition between two phases. One of these
corresponds to {\em one Dirac sea of spinons} (at small $\theta$), while the
other one is connected with {\em two Dirac seas for the same kind of spinons}
(at large $\theta$). It can also be seen from the r.h.s. of Eqs.~(\ref{rho}),
(\ref{eps}) for the densities and dressed energies that the bare density and
bare energy (corresponding to terms which do not depend on $\rho (u)$ and
$\varepsilon (u)$) have either {\em one} or {\em two} minima, respectively.
Hence, they reproduce the same property in the dressed characteristics: The
interaction simply ``dresses'' the (quasi)particles, as usual, but the
``dressing'' does not affect the picture qualitatively. The new critical
field value can be approximated by $h_c \approx {\textstyle \frac {\pi}
{\gamma }}\sin \gamma \cosh^{-1}{\textstyle \frac {\pi \theta}{\gamma}}$ in 
the first order approximation \cite{Zv3}. In this approximation the 
tricritical point is the root of the equation $1 \approx \sinh {\textstyle 
\frac {\pi \theta_c}{\gamma}}$. At this point two second order phase 
transition lines $h_s$ and $h_c$ join. Hence, the ``easy-plane'' magnetic 
anisotropy in the antiferromagnetic two-chain model {\em does not} change 
qualitatively the groundstate behavior in the external magnetic field, cf. 
\cite{Zv3,FR}. However it changes the critical values of the magnetic field 
and the intra-chain coupling. The difference between the two (gapless) phases 
is obvious: the first phase corresponds to the N\'eel-like antiferromagnetic
groundstate for spins in {\em both} chains (along the zigzag line), while the 
second phase is connected with the N\'eel-like antiferromagnetic
groundstates in {\em each of chains}, i.e. to effectively two magnetic
sublattices in the two-chain model.

That is why our simple model explains in which domains of parameters the
two-chain spin system behaves like one-sublattice quantum ``easy-plane''
antiferromagnet, and where it behaves as the two-sublattice one. Note also
that the phase transitions we study here are the manifestations of the
commensurate-incommensurate phase transitions for spin systems. One can
obviously see this, because the {\em intra-chain coupling for two spin chains}
can be interpreted as the {\em next-nearest neighbor spin interactions for a 
single spin chain} of higher length $N_+ + N_-$. Here the magnetic couplings 
are spin-frustrated, thus the emergence of the incommensurate magnetic states 
is understandable.

As a consequence of the conformal invariance of (1+1)-dimensional quantum
systems, the classification of universality classes is simple in terms of the
central charge (conformal anomaly $C$) of the underlying Virasoro algebra
\cite{cft}. The critical exponents in a conformally invariant theory are
scaling dimensions of the operators within the quantum model. They can be
calculated considering the finite-size (mesoscopic) corrections for the 
energies and quasimomenta of the groundstate and low-lying excited states. 
Conformal invariance formally requires all gapless excitations to have the same
velocity (Lorentz invariance). The complete critical theory for systems with
several gapless excitations with {\em different} Fermi velocities is usually
given as a {\em semidirect product} of these independent Virasoro algebras.
\cite{conf} Here we briefly sketch the procedure and write the results for
the finite-size corrections to the energy, following the standard procedure,
see, e.g. Refs.~\cite{conf}. One can see that for $\theta < \theta_c$
and for $\theta > \theta_c$, $h < h_c$, the conformal limit of our uniaxial
two chain spin ${1\over2}$ model corresponds to {\em one} level-1 Kac-Moody
algebra ({\em one} WZW model of level 1 with the conformal anomaly $C = 1$).
The finite-size correction to the energy is rather standard (cf. \cite{conf})
\begin{equation}
E_{fs} (N_+ + N_-) = - {\pi \over 6}v_F + 2\pi v_F (\Delta_l + \Delta_r)
\ , \
\label{E1}
\end{equation}
where $v_F$ is the Fermi velocity of the spinon and the conformal dimensions
of primary operators are (please, pay attention: the lower indices denote the
conformal dimensions for right- and left-moving quasiparticles, at the right
and left Fermi point, respectively):
\begin{equation}
2\Delta_{l,r} = \biggl( {\Delta M \over 2z} \pm z \Delta D \biggr)^2 +
2n_{l,r} \ , \
\label{d1}
\end{equation}
where $\Delta M$ is an integer denoting the change of the number of particles
induced by the primary operator, $\Delta D$ is an integer (half-integer)
denoting the number of transfered particles from the right to the left Fermi
point (backward scattering processes), $n_{l,r}$ are the numbers of the
particle-hole excitations of right- and left-movers. The values for the
quantum numbers are restricted by $\Delta D = \Delta M / 2$ (mod 1). The
dressed charge $z = \xi (Q)$ is the solution of the (standard) integral
equation \cite{conf}
\begin{equation}
\xi (u) + \int_{(Q)}dv K(u-v) \xi(v) = 1 \ ,
\label{xi1}
\end{equation}
taken at the limits of integration (these are the Fermi points, symmetric
with respect to zero). In this phase there is only one region of integration
over $v$. The dressed charge is a scalar. The behavior of our class of models
in this phase in the conformal limit is rather standard \cite{conf}. The
correlation functions decay asymptotically $\propto (x-v_Ft)^{-\Delta_l}
(x+v_Ft)^{-\Delta_r}$. The choise of the appropriate quantum numbers of
excitations $\Delta M$, $\Delta D$ and $n_{l,r}$ is determined for the
leading asymptotics of correlators by taking the possible numbers with
smallest exponents.

But for $\theta > \theta_c$, $h > h_c$, the conformal limit of the
``easy-plane'' two-chain spin ${1\over2}$ model corresponds to the semidirect
product of {\em two} level-1 Kac-Moody algebras, both with conformal
anomalies $C = 1$, i.e., to {\em two} WZW models both of level 1
\cite{Zv3,FR}. The Dirac seas (i.e. the possible spinons with negative
energies) are in the intervals $[-Q^+,-Q^-]$ and $[Q^-,Q^+]$ (minima in the
distributions of rapidities at $\mp \theta$). This can be interpreted as the
symmetrically distributed (around zero) Dirac seas of {\em ``particles''} for
$[-Q^+,Q^+]$ and the Dirac sea of {\em ``holes''} for $[-Q^-,Q^-]$. In fact the
valley in the density distribution for ``particles'' and the maximum for
``holes'' are in the one-to-one correspondence with the maxima and minimum of
the dispersion law for spinons. The second critical field $h_c$ in this
language corresponds to the van Hove singularity of the empty band of
``holes''. Naturally, the Fermy velocities of ``particles'' are positive,
$v_F^+ = (2\pi \rho(Q^+))^{-1}\varepsilon'(u)|_{u=Q^+}$, while the Fermy
velocities of ``holes'' are negative $v_F^- = -(2\pi \rho(Q^-))^{-1}
\varepsilon'(u)|_{u=Q^-}$. The finite-size corrections to the energy for this
case are
\begin{equation}
E_{fs}(N_+ + N_-) = -{\pi \over 6}(v_F^+ + v_F^-) +
2\pi \biggl(v_F^+ (\Delta_l^+ + \Delta_r^+) +
v_F^-(\Delta_l^- + \Delta_r^-)\biggr) \ , \
\label{E2}
\end{equation}
where the dispersion laws of ``particles'' and ``holes'' are linearized about
the Fermi points for each Dirac sea. The conformal dimensions of the primary
operators are (the upper indices denote Dirac seas; the lower indices denote 
right and left Fermi points of each of these two Dirac seas, cf. \cite{FR} 
for the isotropic spin ${1\over2}$ two-chain model):
\begin{equation}
2\Delta_{l,r}^{\mp} = \biggl[{(x_{-\pm}\Delta M^+ -
x_{+\pm}\Delta M^- )\over 2\det {\hat x}} \mp
{(z_{-\pm}\Delta D^+ - z_{+\pm}\Delta D^-) \over 2\det
{\hat z}}\biggr]^2 + 2n_{l,r}^{\mp} \ ,
\label{d2}
\end{equation}
where the ``minus'' sign between the terms in square brackets corresponds to
the right-, and ``plus'' sign to the left-movers. Here $\Delta M^{\pm}$ denote
the differences between the numbers of particles excited in the Dirac seas of
``particles'' and ``holes'', labeled by the upper indices. $\Delta D^{\pm}$
denote the numbers of backward scattering excitations, and $n_{l,r}^{\pm}$
are the numbers of the particle-hole excitations for right- and left-movers
of each of Dirac seas (for ``particles'' and ``holes''). Please pay attention 
that $\Delta M^{\pm}$ and $\Delta D^{\pm}$ are not independent. Their values 
are restricted by the following connections: $\Delta M^+ - \Delta M^- = 
\Delta M$, and $\Delta D^+ - \Delta D^- = \Delta D$, where $\Delta M$ and 
$\Delta D$ determine in a standard way the changes of the total magnetization 
and the total momentum of the system, respectively, due to excitations. 
Please note that in Refs.~\cite{FR,FR1} these restrictions were missing; this 
resulted in, e.g., the invalid statement that {\em four independent} 
backscattering low-lying excitations are possible. However one can see that 
only two of them are really independent. The same is true for the 
excitations that change the total magnetization of the system: there are only 
{\em two independent} of four possible such excitations. This is a direct 
consequence of the fact that {\em only one magnetic field} determines the 
filling of the Dirac seas for ``particles'' and ``holes'', or, in other 
words, two Dirac seas for spinons at $\pm \theta$.

The dressed charges $x_{ik}(Q^k)$ and $z_{ik}(Q^k)$ ($i,k =+,-$) are matrices
in this phase. They can be expressed by using the solution of the integral
equation \cite{Kor,conf}
\begin{equation}
f(u|Q^{\pm}) = \biggl(\int_{-Q^+}^{Q^+} - \int_{-Q^-}^{Q^-}\biggr) K(u-v)
f(v|Q^{\pm}) =
K(u-Q^{\pm}) \ ,
\label{f}
\end{equation}
with \cite{conf}
\begin{eqnarray}
&&z_{ik}(Q^k) = \delta_{i,k} + (-)^k {1\over2}\bigl(
\int_{Q^i}^{\infty} - \int_{-\infty}^{-Q^i}\bigr)dv f(v|Q^k)
\nonumber \\
&&x_{ik}(Q^k) = \delta_{i,k} - (-)^k \int_{-Q^i}^{Q^i}dv f(v|Q^k) \ .
\label{xi2}
\end{eqnarray}
Notice, please, that the dressed charges depend on the value of the magnetic
anisotropy $\gamma$ via the kernels, while they depend indirectly on the
value of the intra-chain coupling constant $\theta$, only via the limits of
integrations. In the first order approximation one can write the solutions as
$x_{ik}(Q^k) \approx \delta_{i,k} - (-)^k \int_{-Q^i}^{Q^i}dv K(v - Q^k) +
\dots$ and $z_{ik}(Q^k) \approx \delta_{i,k} + (-)^k (1/2)
(\int_{Q^i}^{\infty} - \int_{-\infty}^{-Q^i})dv K(u - Q^k) + \dots$. The
Dirac sea for ``holes'' disappears, naturally for $h \to h_c$, $\theta \to
\theta_c$. The slopes of the dressed energies of ``particles'' and ``holes''
at Fermi points of the Dirac seas (Fermi velocities) differ in general from
each other. Therefore we have a semidirect product of two algebras. Hence, in
this region the dressed charges are $2\times2$ matrices. This means that the
conformal limit of the ``easy-plane'' two-chain spin ${1\over2}$ model
corresponds to {\em one} or {\em two} WZW theories, depending on the values of
the intra-chain coupling, magnetic anisotropy and magnetic field. At the
critical line $h_c$ the Dirac sea of ``holes'' disappears as well as the
components of the dressed charge matrix $\hat x$ (with square root
singularities of the critical exponents for the correlation functions). Note 
that the dressed charge $z$ becomes $z=(2x)^{-1}$ at the phase transition line
$h_c$. This corresponds to the disappearence of one of the WZW CFTs.
Unfortunately it is impossible to obtain an analytic solution to
Eqs.~(\ref{f}) in closed form for a finite inter-chain coupling $\theta$.
Naturally in the limiting cases of two independent chains of lengths
$N_{\pm}$, $\theta \to \infty$, and a single chain of length $N_+ + N_-$,
$\theta = 0$, the solutions of Eqs.~(\ref{xi1}),(\ref{f}),(\ref{xi2})
coincide with well-known ones, see Refs.~\cite{conf}. The correlation
functions of the uniaxial two-chain spin ${1\over2}$ model decay
algebraically in this phase $\propto (x-v_F^+t)^{-\Delta_l^+}
(x+v_F^-t)^{-\Delta_l^-}(x-v_F^+t)^{-\Delta_r^+}(x+v_F^-t)^{-\Delta_r^-}$
with the minimal exponents of possible quantum numbers of excitations
$\Delta M^{\pm}$, $\Delta D^{\pm}$ and $n_{l,r}^{\pm}$. We point out once
more that the same magnetic field plays the role of a chemical potential for
the ``particles'' and ``holes'', or spinons of both Dirac seas in the second
phase, and hence this choise of ``minimal quantum numbers'' is constrained.

We must point out here that there is a crucial difference between our
situation and the case of dressed charge matrices appearing for systems with
the internal structure of bare particles \cite{conf}. There the two Dirac 
seas of the groundstates are connected with different kinds of excitations, 
e.g., holons and spinons for the repulsive Hubbard model, or Cooper-like 
singlet pairs and spinons for the supersymmetric $t\!-\!J$ model. They 
correspond to two different kinds of Lagrange multipliers, the chemical 
potentials and magnetic fields. Thus the low-lying excitations of the 
conformal theories in the spin and charge sectors of these correlated 
electron models are practically {\em independent} of each other (spin-charge 
separation). Note that the spin and charge sectors are connected via the 
off-diagonal elements of the dressed charge matrix though. This is the 
consequence of the fact that say, holons or unbound electrons carry {\em both 
charge and spin}. On the other hand, two Dirac seas appear for the {\em same 
kinds of particles} for the models studied in this paper, which are also 
connected with the {\em same} magnetic field governing the filling of both 
Dirac seas. The latters appear due to two minima in the bare energy 
distribution and correspond to nonzero shift $\theta$ in the Bethe ansatz 
equations. In other words, two Dirac seas are determined by the inter-chain 
coupling and appear if the values of coupling and external magnetic field are 
higher than the {\em threshold} values $\theta_c$ and $h_c$, respectively. We 
believe that such a threshold behavior does not depend on the integrability 
of the model and is the generic feature for any multi-chain quantum spin 
models.

The low temperature Sommerfeld approximation shows that as usually the low
temperature specific heat is proportional to $T$ out of critical lines. At
the critical lines the van Hove singularities produce $\sqrt{T}$ low
temperature behavior of the specific heat, while at the tricritical point we
have $T^{1/4}$ behavior.

What are the changes due to the different lengths of the chains $N_+ \ne N_-$?
One can see obviously that the values of the momentum, energy and velocity of
a spinon (which was $v = (\pi/ \gamma)\sin \gamma \tanh (\pi \theta/
\gamma)$) become functions of $N_+ - N_-$. For example, the velocity
renormalizes as $v \to v[1 + (N_+ - N_-)^2 \tanh^2(\pi \theta/ 2\gamma)
/N^2]^{-1}$. This introduces dependences of the critical values $\theta_c$ 
and $h_c$ (as well as the saturation field $h_s$) on the difference $N_+ - 
N_-$. Also, the Fermi velocities and Fermi points for the finite-size 
corrections become dependent on this difference. One can in principle 
consider different coupling constants $J_{\pm}$ for each of the chains 
(overall multipliers \cite{PL}). This produces the renormalizations
similar to the action of $N_+ \ne N_-$, i.e., the velocity, e.g., renormalizes
as $v \to J_+v[1 + (J_-/J_+)^2 \tanh^2(\pi \theta/ 2 \gamma)]^{-1}$.

\section{\bf Connections to the Quantum Field Theories}

The studies presented in the previos section, being rather standard (note,
though, some important new features, which were absent in the previous studies
\cite{PopZv,Zv1,Zv2,Zv3,FR,FR1}, such as the dependencies of the critical 
values of the inter-chain coupling and external magnetic field on the 
parameter of the magnetic anisotropy and on the difference in the lengths of 
the chains; also the important restrictions on the quantum numbers of 
low-lying conformal excitations). However we will use the results of that 
section for novel studies for a wider classes of exactly solvable models in 
Sections~3-5. For instance, in this section we point out the important 
similarities in the behaviors of the two-chain quantum spin model considered 
in the previous section and several models of QFT.

Really, when examinating Eqs.~(\ref{BAE}), one can see that these Bethe
ansatz equations coincide with the ones, which describe the behavior of the
spin (color) sector of some QFT. $N_{\pm}$ corresponds to the numbers of
(bare) particles with the positive and negative chiralities. For example,
for the chiral-invariant Gross-Neveu model \cite{AL1,JN} we have to put
$\gamma \to 0$ (i.e. $SU(2)$-symmetric case, equivalent to the
$SU(2)$-symmetric Thirring model), and $\theta = (1 - g^2)/2g$, where $g$ is
the coupling constant of the chiral invariant Gross-Neveu QFT \cite{AL1}. As
for the Lagrange multiplier $h$, it can play the roles of either an external
magnetic field, or the chemical potential, or an external topological field,
dual to the topological Noether current in QFT. Here we point out that in
fact in QFT theorists are interested in physical particles, which have the
finite mass (gap). In the chiral-invariant Gross-Neveu model the gap of the
staggered oscillations of the two-chain quantum spin model plays the role of
the physical mass of the particle (spinor) \cite{AL1,Zv2}. As for the
(gapless) oscillations of the magnetization of the two-chain spin model, we
point out that they are the consequences of the lattice, and play the role of
the massless fermion doublers of the lattice QFT \cite{DS}. The results of
the previous section mean that the behavior of the chiral-invariant
Gross-Neveu model (or $SU(2)$-symmetric Thirring model) in an external
magnetic field depends strongly on the coupling constant $\theta$ (or
equivalently on $g$). For $\theta < \theta_c$ the conformal limit of the QFT
corresponds to one level-1 WZW model with the conformal dimension $C=1$.
However for $\theta > \theta_c$ ($-\theta_c - \sqrt{\theta_c^2 +4} < 2g <
-\theta_c + \sqrt{\theta_c^2 +4}$) the conformal limit of this QFT in an
external magnetic field corresponds to the semidirect product of {\em two}
level-1 WZW model with the conformal dimensions $C=1$. Two kinds of conformal
points for this QFT have been mentioned already \cite{DdV2} in a slightly
different context. They were connected with one WZW theory or two WZW
theories, coupled via a current-current interaction. This is related to
right-left symmetry of the chiral invariant Gross-Neveu QFT (see, also,
Refs.~\cite{PW1,PW2} for the case of the QFT for the principal chiral
field).

Note, that the condition $h > h_c$ in the QFT means that the magnetic field 
is larger than the mass of the physical particle (color spinor). In this 
sense, in the region of magnetic field values $h < h_c$ the results of the 
QFT (see, e.g., \cite{JN}) predict zero magnetization, however a different 
lattice regularization, similar to the lattice scheme used in the previous 
section predicts a {\em nonzero} magnetization of the chiral-invariant 
Gross-Neveu model in this region. This is the indirect effect of the fermion 
doublers. In other words, it is connected with the well-known mapping of the 
lattice (e.g., Thirring) model under regularization on two continuum QFTs 
either both {\em bosonic} (free bosonic QFT and sine-Gordon one, \cite{LE}), 
or both {\em fermionic} ones (a free one and the continuum massive Thirring 
model). There are necessarily two such theories because of the 
Nielsen-Ninomiya fermion doublers: remember that we have started from a 
lattice \cite{DS}.

For other models of QFT the procedure of the lattice regularization
\cite{TTF,BI,DdV1} has been used. Here $\theta$ plays the role of the cut-off 
to preserve the mass of the physical particle to be finite. For example, for 
the $U(1)$-symmetric Thirring QFT \cite{JNW,DS} one can use the results of the
previuos section with the limit $\theta \to \infty$ taken {\em after} the
thermodynamic limit ($L,N_{\pm},M \to \infty$ with their ratios fixed, $L$ is
the size of the box). In this case one can obviously obtain the conformal
limit of the theory with nonzero physical masses of the particles. Naturally
in the limit $\theta \to \infty$ we ever exist in an external magnetic field
in the phase with {\em two} Dirac seas. Here the latters correspond to the
right- and left-moving particles (with positive and negative chiralities).
Actually here our point of view coincides with the one of the field
theorists. Recently it was shown in Ref.~\cite{DdV3} that for
(1+1)-dimensional sine-Gordon model the lattice regularization scheme in the
``light-cone'' approach gives similar to ours results for the conformal limit
of the model. It was shown there that at the UV fixed point the conformal
dimensions of the sine-Gordon model are determined by {\em two} $U(1)$
charges of excitations (the usual one and the {\em chiral charge}). The
chiral charge corresponds to the number of excitations transfered from one
Dirac sea to the other, similar to our results (note that the above-mentioned
lattice-regularized sine-Gordon case corresponds in our notations to $\theta
\to \infty$, where the integral equations for the particles with the positive
and negative chiralities are totally decoupled). We point out here, that such
a behavior is not unexpected, because the sine-Gordon QFT belongs to the same
class of models, which is studied in our paper, i.e., its Bethe ansatz
description features a shift of rapidities in the Bethe ansatz equations in
the lattice-regularized theory \cite{DdV3}.

\section{\bf Higher spin (chirality) generalizations}

For the higher spin generalizations of the Bethe ansatz theory presented in
Section~2 we can write down BAE in the form
\begin{equation}
\prod_{\pm}e_{n_{\pm}}^{N_{\pm}}(u_{\alpha}\pm \theta) =
e^{i\pi M}\prod_{\beta=1,\beta \ne \alpha}^M e_2 (u_{\alpha} - u_{\beta}) \ ,
\label{BAE1}
\end{equation}
where $n_{\pm}=2S_{\pm}$ are the values of spins in each chain or the colors
of bare particles in QFT. The eigenvalue of the transfer matrix can be writen 
as
\begin{eqnarray}
&&\Lambda (\lambda) = \prod_{\alpha =1}^M {{\sinh (\lambda - u_{\alpha} +
i\gamma {\textstyle \frac {1}{2}})}\over \sinh (u_{\alpha} - \lambda +
i\gamma {\textstyle \frac {1}{2}})} +
e^{i\pi M} \prod_{\pm}\bigl( {{\sinh (\lambda \pm \theta)}\over \sinh
(i\gamma {\textstyle \frac {n_{\pm}}{2}} - \lambda \mp \theta)} \bigr)^
{N_{\pm}} \times
\nonumber \\
&&\prod_{\alpha =1}^{M} {{\sinh (u_{\alpha}  - \lambda +
i\gamma {\textstyle \frac {3}{2}})}\over \sinh (\lambda - u_{\alpha} -
i\gamma {\textstyle \frac {1}{2}})} \ . \
\label{tran}
\end{eqnarray}

Similar new phases with one or two kinds of Dirac seas for similar kinds of
low-lying excitations exist also for a number of models in which
$n_{\pm} \ne 1$, e.g. for the higher-spin ($S > {1\over2}$) two-chain models
with equal spins in each chain, $SU(n+1)$ CIGN QFT \cite{AL2}: there $n_+ =
n_- = n \neq 1$; for the principal chiral field models (nonlinear
$\sigma$-model) for $CP$-symmetric \cite{PW1} (there $n_+ = n_- \to \infty$)
and $CP$-asymmetric cases \cite{PW2} (there $n_+ \ne n_-$, $(n_+ + n_-) \to
\infty$, $(n_+ - n_-)$ fixed, i.e., the symmetry $SU(2)\times SU(2) \propto
O(4)$); and for the $O(3)$-symmetric nonlinear $\sigma$-model \cite{Wie2} as
well as for spin-$(S_+ \equiv 2n_+)$ - spin-$(S_- \equiv 2n_-)$ two-chain
models ({\em quantum two-chain ferrimagnet}). Note that for spins $S \ne
{1\over2}$ the procedure of the construction of the Hamiltonian is more
complicated, because it corresponds to the two-chain uniaxial generalization
of the Takhtajan-Babujian model, see e.g., Refs.~\cite{TaBa}. For the
simplest case of the isotropic exchange interaction between the spins and
between the chains the Hamiltonian has the form:
\begin{eqnarray}
&&{\cal H} = \sum_n \biggl( \theta^2({\cal H}_{S_+,S_+,n_1,n_1+1} +
{\cal H}_{S_-,S_-,n_2,n_2+1}) +
2({\cal H}_{S_+,S_-,n_1,n_2} + {\cal H}_{S_+,S_-,n_1,n_2+1}) +
\nonumber \\
&&\{({\cal H}_{S_+,S_+,n_1,n_1+1} + {\cal H}_{S_-,S_-,n_2,n_2+1}),
({\cal H}_{S_+,S_-,n_1,n_2} + {\cal H}_{S_+,S_-,n_1,n_2+1})\} +
\nonumber \\
&&2i\theta \bigl[({\cal H}_{S_+,S_+,n_1,n_1+1} +
{\cal H}_{S_-,S_-,n_2,n_2+1}),
({\cal H}_{S_+,S_-,n_1,n_2} + {\cal H}_{S_+,S_-,n_1,n_2+1}) \bigr] \biggr)
\ ,
\label{Hss}
\end{eqnarray}
where $[.,.]$ ($\{.,.\}$) denote (anti)commutator,
\begin{equation}
{\cal H}_{S_1,S_2,n,n+1} = \sum_{j=|S_1-S_2|+1}^{S_1+S_2}
\sum_{k=|S_1-S_2|+1}^j {k\over k^2 + \theta^2} \times
\prod_{l=|S_1-S_2|}^{S_1+S_2}{x-x_l\over x_j -x_l} \ , \
\label{hnss}
\end{equation}
$x = {\vec S}_{1,n}{\vec S}_{2,n+1}$, and $2x_j = j(j+1) - S_1(S_1+1) -
S_2(S_2+1)$. The summation over $n$ runs to $N_{\pm}$ in each chain. One can
obviously see that for $S_{\pm} = {1\over2}$ the Hamiltonian Eq.~(\ref{Hss})
recovers the isotropic antiferromagnetic spin ${1\over2}$ Hamiltonian
Eq.~(\ref{isHam}) investigated in Section~2. For a single spin chain,
$\theta = 0$, $N_+=N_-$ the Hamiltonian coincides with the known one of
alternating spin chains \cite{dVW,Sch,ZS}. The Bethe ansatz studies of
the model for $n_{\pm}$ can be performed in the complete analogy with the above
mentioned case $n_{\pm} = 1$, keeping in mind, of course, the main difference:
for the $SU(2)$-symmetric or uniaxial higher spin models the groundstate
corresponds to the filling up the Dirac seas for spin strings of lengths
$n_{\pm}$. \cite{TaBa} The well-known fusion scheme can be used for the case
of a flavor-degenerate situation of the chiral invariant Gross-Neveu QFT, in
the absence of flavor fields \cite{KRS}. Note that, except of the
$O(3)$-symmetric case, $\gamma = 0$ everywhere in the above-mentioned models 
of QFT. This corresponds to rational solutions of the Yang-Baxter equation 
for the two-particle scattering matrices. For the two spin chains the two-chain
quantum ferrimagnet model corresponds to {\em two} Takhtajian-Babujian chains
of different values of site spins coupled due to nonzero $\theta$. The total
quasimomentum and the energy of the system in the framework of the lattice
(local) regularization scheme for some QFT can be written as \cite{DS}
\begin{eqnarray}
&&- 2a_t E = \sum_{\pm} \sum_{\alpha = 1}^M {{\partial}\over \partial
u_{\alpha}} N_{\pm} \ln e_{n_{\pm}}(u_{\alpha} \pm \theta )
\nonumber \\
&&i a P = \sum_{\pm} \sum_{\alpha = 1}^M N_{\pm} \ln e_{n_{\pm}}(u_{\alpha}
\pm \theta)  \ ,
\label{EP}
\end{eqnarray}
where $a$ and $a_t$ denote the space and time lattice constants, respectively,
and their ratio fixes the velocity of light (``light-cone'' approach). The
$CP$-symmetric (chiral invariant) case corresponds to the situation, in which
$n_+ = n_- = n$. The Dirac seas are related to the dressed (quasi)particles
with negative energies (strings of length $n_{\pm}$). The behavior of the
dispersion law for excited particle in the $CP$-symmetric case ($n_+ = n_- =
n$ and $N_+=N_-$) is similar to Eq.~(\ref{disp}): for instance, for the
chiral-invariant Gross-Neveu QFT and principal chiral field model the r.h.s.
of Eq.~(\ref{disp}) must be simply multiplied by $\sin (\pi r /n+1)/ \sin
(\pi / n+1)$ and the parameter $\theta$ in Eq.~(\ref{disp}) has to be
replaced by $(n+1) \theta /2$. $r = 1, \dots , n$ is the rank of a fundamental
representation of the $SU(n+1)$ algebra. All the previously mentioned
characteristic features from the case $n_{\pm} =1$ persist. The differences
are in the levels of Kac-Moody algebras in the conformal limit: The
conformal anomalies are $C = \textstyle \frac {3n}{n + 2}$. Now the
conformal field theory is a semidirect product of a Gaussian ($C=1$)
\cite{KB} and $Z(n)$ parafermion models \cite{par1}: the operators identified
from the scaling behavior of states consisting of Dirac sea strings only
(found from finite-size corrections) are found to be composite operators
formed by the product of a Gaussian-type operator and the operator in the
parafermionic sector. To find a nonzero contributions from parafermions
(constant shifts) one can consider the states with strings of other lengths
then the Dirac sea present \cite{par2}. For the scaling dimensions these
shifts are ${2r -r^2\over2n+4}$, $r = 1,2,\dots$.

From now on we concentrate on the $n_+ \ne n_-$ situation. For the two-chain
spin system the situation corresponds to the quantum ferrimagnet. Here we
point out that due to the zigzag-like interactions in the system and spin
frustration the ferrimagnets of this class are in the {\em singlet}
groundstate (compensated phase) for $h=0$, unlike the standard classical
ferrimagnets in uncompensated phases. The integral equations that determine
the physical vacuum of the systems are similar to
Eqs.~(\ref{rho})-(\ref{eps}). They reveal one or several minima of the
corresponding distributions of dressed energies and densities with possible
negative energy states, i.e., one or several Dirac seas:
\begin{eqnarray}
&&\varepsilon_{\tau}(u) + \int dv K_{\tau \tau'}(u-v)
\varepsilon_{\tau'}(v) = h {\textstyle \frac {N_{\tau}}{N}}n_{\tau} +
\sum_{\pm}{\textstyle \frac {N_{\pm}}{N}}\varepsilon_{\tau,\pm}^0
\nonumber \\
&&\rho_{\tau}(u) + \int dv K_{\tau \tau'}(u-v) \rho_{\tau'}(v) =
\sum_{\pm}{\textstyle \frac {N_{\pm}}{N}}
\rho_{\tau,\pm}^0 \ .
\label{eps2}
\end{eqnarray}
The index $\tau$ enumerates two possible Dirac seas and appears due to $n_+
\ne n_-$, and the $\pm$ enumerate two possible minima due to the nonzero shift
$\theta$. The index $\tau$ was naturally absent for the $CP$-symmetric case
$n_+ = n_-$. Note that for quantum two-chain ferrimagnets the investigated 
gapless phases in the groundstate in an external magnetic field are similar 
to the spin-compensated and uncompensated phases. Thus the phase transition 
between those phases is similar in nature to the well-known spin-flop phase 
transition in the classical theory of magnetism. Note, though, that the 
spin-flop transition is of the first order (``easy-axis'' magnetic 
anisotropy), while the transition under study is the second-order one 
(``easy-plane'' anisotropy). The Fourier transform of the kernel is given by
\begin{eqnarray}
&&2\coth (\omega /2) \bigl[ diag \bigl(e^{-n_+|\omega /2|}\cosh
(n_+\omega/2), e^{-n_-|\omega /2|}\cosh (n_-\omega /2) \bigr) +
\nonumber \\
&&{\hat \sigma}_{x}\bigl(e^{-(n_+ - n_-)|\omega /2|} -
e^{-(n_+ +n_-)|\omega / 2|}\bigr)\bigr] \ .
\label{kern}
\end{eqnarray}
$diag(a,b)$ is $2\times2$ diagonal matrix and ${\hat \sigma}_x$ is the usual
Pauli matrix. Note, that after taking the limit $(n_+ + n_-) \to \infty$,
which is the case of the $CP$-asymmetric case of the QFT for the principal
chiral field, i.e. with the Wess-Zumino term \cite{PW2}, the inverse kernel
coincides formally (up to a constant multiplier) with the one for the case
$n_+ = n_- = 1$. This indicates the global $O(4)$ ($O(3)$) symmetry of the
principal chiral field \cite{PW2}. There may be also two different behaviors,
corresponding to one or several Dirac seas for $n_+ \ne 1$ or $n_- \ne 1$.
Naturally in the conformal limit the associated WZW CFTs have different
conformal anomalies determined by $n_{\pm}$: $C_{\pm} = \textstyle \frac
{3n_{\pm}}{n_{\pm} + 2}$. For the determination of the Gaussian parts of the
conformal dimensions of primary operators Eqs.~(\ref{d2}) can be used. One
has to add the input from the parafermionic sectors, too \cite{par1,par2}.
The elements of the dressed charge matrices are the solutions of the
following system of integral equations:
\begin{equation}
\xi_{\tau,\tau'}(u) + \sum_{\pm} \int dv K_{\tau'}(u-v) \xi_{\tau,\pm}(v) =
\delta_{\tau,\tau'} \ ,
\label{xi3}
\end{equation}
in which the summation over $\pm$ is due to the two possible Dirac seas (two
minima in the distribution of rapidities) at $\pm \theta$. For different 
values of the spins, $n_+ \ne n_-$, a transition between two different phases 
is induced by increasing an external magnetic field to some critical value,
even in the absence of the shift $\theta$ \cite{Sch,ZS}. It differs from the
$CP$-symmetric case $n_+ = n_-$, where the phase transition is only connected
with the nonzero value of the intra-chain coupling parameter $\theta$. For
the $CP$-symmetric case one or two Dirac seas of the same type of excitations
exist due to nonzero $\theta$. But in the $CP$-asymmetric case the existence
of two Dirac seas can be related to two kinds of different low-lying
excitations (particles). They are strings of lengths $n_+$ and $n_-$,
respectively. In this situation the dispersion laws may be independent (not
only factorized as for the previous $CP$-symmetric cases). The (new) phase
transition at $h_c$ reveals the van Hove singularity of the empty Dirac sea
for the longer strings. The spin saturation field $h_s$ is connected with the
empty Dirac sea of strings of the smaller length.

\section{\bf Multi-chain quantum spin models}

It is worthwile to mention that phase transitions in an external magnetic
field, similar to the ones studied in this paper for uniaxial spin chains
and QFT, have been already studied in the 1D quantum alternating single spin
chains \cite{Sch,ZS}, spin ${1\over2}$ isotropic two-chain models
\cite{Zv3,FR}, and correlated electron models with the finite concentration of
magnetic impurities \cite{SchZv}. The Bethe ansatz equations for those models
are similar to the ones studied in the present paper,
Eqs.~(\ref{BAE}),(\ref{BAE1}). Note that the energies for spin models are
defined (as usual for the lattice models) as first logarithmic derivatives of
the transfer matrices. The factorization of the dispersion law for the lowest
excitations (spinon) reveals essentially {\em two} kinds of magnetic
oscillations: excitations of the magnetization and oscillations of the
staggered magnetization, i.e., the manifestation of {\em essentially two
magnetic sublattices}. Naturally, the existence of the latters persists in the
continuum limit of such systems too (cf., for instance, with the standard
theory of antiferromagnetism). Two non-ferromagnetic phases also reveal
themselves in finite-size corrections to the energies of these quantum spin
models. There instead of a scalar dressed charge for the phase with one Dirac
sea for spinons, $2\times 2$ dressed charge matrices appear in the second
phase with two Dirac seas for for spin strings of different lengths in
alternating spin chain \cite{Sch,ZS}, or for spinons of the same kind in
zigzag-like coupled spin chains (see \cite{Zv3,FR} for the isotropic
two-chain spin-${1\over2}$ model).

The symmetry-breaking terms [the difference $(n_+ - n_-) = 2(S_1 - S_2)$, or
nonzero $\theta$] in BAE are actually the reason for the emergence of
several gapless phases (or two Dirac seas) in the groundstate in an external
magnetic field. It is also interesting to note that a homogenuous shift of
rapidities can be removed for one Dirac sea for the {\em periodic} boundary
conditions by an appropriate unitary (gauge) transformation (shift of
variables), e.g., $u_{\alpha} \to u_{\alpha} \pm \theta$. But in the case of
{\em open} boundary conditions, BAE take the form (for simlicity reasons we
write the {\em free} boundary situation only, without the external boundary
potential):
\begin{equation}
\prod_{\pm} e_{n_{\pm}}^{2N}(u_{\alpha} \pm \theta ) =
\prod_{\pm} \prod_{\beta}e_2(u_{\alpha} \pm u_{\beta}) \ .
\label{oBAE}
\end{equation}
It is clear that for the open chain one cannot remove the shift $\theta$ of
rapidities $u_{\alpha}$ from one Dirac sea by a special choice of the gauge.
From this point of view the latter case is close to the $CP$-asymmetric
situation in QFT.

One can see from the structure of the Hamiltonians that for the two-chain
spin models the parameter $\theta$ characterizes the intra-chain coupling for
each chain (or the next-nearest-neighbor interaction in a single spin
chain picture). It is obvious to introduce the series of
$\{\theta_j\}_{j=1}^J$ (for each chain) and to construct the Hamiltonian of
the exactly integrable multi-chain ($J$ is the number of chains) spin model.
For the simplest case of all $S={1\over2}$ isotropic antiferromagnetic chains
the Hamiltonian reads \cite{PopZv}:
\begin{eqnarray}
&&{\hat H}_J = A \sum_n \Bigl( \biggl( \prod_{i,k}(\theta_i -
\theta_k) \biggr) {\hat P}_{S_{n,r}S_{n+1,r}} + \sum_{p<q}
{\prod_{i,k} (\theta_i -\theta_k) \over (\theta_p - \theta_q)}
[{\hat P}_{S_{n,q}S_{n+1,p}},{\hat P}_{S_{n,q}S_{n+1,q}} +
\nonumber \\
&&{\hat P}_{S_{n,p}S_{n+1,p}}] + \ldots +
\biggl( \sum_{j=1}^J
{\hat P}_{S_{n,j}S_{n,j+1}} -
{\hat P}_{S_{n,J}S_{n,J+1}} + {\hat P}_{S_{n,J}S_{n+1,1}} \biggr)
\Bigr) \ ,
\end{eqnarray}
where $A$ is the normalization constant (which depends on $\theta_j$),
${\hat P}_{S_aS_b} \propto (1/2){\hat I}\otimes {\hat I} + 2{\vec S}_a
\otimes {\vec S}_b$ is the permutation operator and $[.,.]$ denotes a
commutator. Note, that in the case of $J \neq 2$ the integrable model
corresponds to the pair couplings not only between the nearest-neighbor 
spins but also to the next-nearest three spins, etc., couplings. All those
terms are only essential in quantum mechanics, because in classical physics 
they are total time derivatives \cite{Zv1} and do not change equations of 
motion. The Bethe ansatz equations have the form:
\begin{equation}
\prod_{j=1}^J e_1^{N_j}(u_m + \theta_j - \theta_1) = e^{i\pi M}\prod_k^M
e_2 (u_m - u_k) \ ,
\end{equation}
where $M$ is the total number of down spins and $N_j$ is the number of sites
in the $j$-th chain. The previously studied situation $J=2$ corresponds to 
the shift of the variables $u_m \to u_m + \theta$ with $\theta_2 - \theta_1 = 
- 2\theta$. Now $\theta_j - \theta_1$ determines the values of the 
intra-chain couplings in chain $j$.

The analysis of the low-temperature thermodynamics of the multi-chain spin
system is analogous to the situation of $J=2$ studied in Sections~2-4. From 
the structure of the Bethe ansatz equations in the thermodynamic limit 
$N_j,M \to \infty$, their ratios fixed, one can see that for $J$-chain model
(for different $\theta_j$) there can exist, generally speaking, $J$ phase
transitions of the second order in the groundstate in an external magnetic
field. These are nothing else than the commensurate-incommensurate phase
transitions for the quantum multi-chain spin model with different couplings
between the chains. The values of the critical fields $h_{c_1}, \ldots,
h_{c_{J-1}}$ and the value of the magnetic field of the transition to the
ferromagnetic state $h_s$ depend on the set of $\theta_j$, i.e., on the
intra-chain couplings (and also on the values of the magnetic anisotropy
constants, which can be taken different for each chain; this does not destroy
the integrability). The ferromagnetic state is gapped, while all other phases 
are gapless in the integrable multi-chain spin quantum model. There are also 
$J-1$ tricritical points at which the lines of the phase transitions 
$h_{c_j}$ join the line of the spin-saturation phase transition. Naturally, 
the phase that corresponds to the lowest value of the magnetic field, say 
$h < h_{c_1}$ for special values of $\theta_j$ (the condition is similar to 
$\theta < \theta_c$ for $J=2$), has in the conformal limit one scalar dressed 
charge. Hence, in the conformal limit our multi-chain spin model behaves as 
the level-1 WZW CFT. In the next phase the multi-chain quantum spin model 
behaves as the semidirect product of two WZW CFTs, hence their dressed 
charges are $2\times 2$ matrices, and so on, until the last gapless phase, 
which corresponds to the semidirect product of $J$ WZW CFTs with $J\times J$ 
dressed charge matrices. Note that $J$ in this approach also denotes the 
number of possible Dirac seas (each of them is connected with the same 
magnetic field, so the excitations in each of them are not independent), and, 
thus, with one-half of the number of Fermi points. In the limit $J \to 
\infty$ (i.e. quasi-2D spin system) one obtains the (2D) Fermi surface 
instead of the set of 1D Fermi points (the latters become disributed more
closely to each other with the grouth of $J$). In this limit the differences
between $\theta_j$ tend to zero, and that is why the differences between
$\theta_{c_j}$, $h_{c_j}$ and also between $h_{c_j}$ and $h_s$ disappear,
too. Therefore in this limit the only $h_s$ survives. It means that for the
quasi-2D limit of such an integrable model of $J$ coupled quantum spin chains
for $J \to \infty$ we expect only {\em two phases} in the groundstate in an
external magnetic field: the ferromagnetic gapped one and the gapless phase,
which in the conformal limit corresponds to {\em one} WZW CFT (with single
{\em scalar dressed charge}). The phase transition between these two phases
in the groundstate in an external magnetic field is of the second order.

\section{Behavior of the non-integrable multi-chain spin systems}

So far we have studied only {\em integrable} multi-chain quantum spin models.
We have shown that the commensurate-incommensurate phase transitions of the
second order have to reveal themselves in an external magnetic field due to
the intra-chain interactions (or the next-nearest interactions in a single
quantum spin chain picture). We have shown that the emergence of these phase
transitions does not depend on the value of the site spins, they emerge in the
presence of the ``easy-plane'' magnetic anisotropy, which keeps the system in
the critical (gapless) region. It is not clear, however, which features of
the behavior of the integrable models with the ``fine-tuned'' parameters have
to exist for more realistic multi-chain models, and what are the qualitative
differences, we expect to exist between the integrable multi-chain models and
real multi-chain spin systems.

We have to add one more thing to clarify the situation: We study (quasi) 1D
spin quantum models, for which one can use the Lieb-Schultz-Mattis theorem
(and its generalizations) \cite{LSM,OYA}. However, it is obvious that due to
the frustration of the interactions between neighboring spins, and the 
presence of additional terms in the Hamiltonians, which violate the 
time-reversal and parity symmetries in the systems (spin chiralities or spin 
currents), for all spin models studied in the paper one cannot satisfy the 
conditions of the theorem. Hence it cannot be applied (at least directly). 
That is why for all the models we study there are no spin gaps (except for 
the trivial one for the spin-polarized groundstate). (Here we are not talking 
about the gaps connected with the magnetic anisotropy, but rather about the 
Haldane-like spin gaps \cite{Hal} which appear even for the isotropic spin-spin
interaction, and about fractional magnetization plateaux \cite{OYA}). As we
argued before \cite{Zv1}, namely the presence of the chiral spin terms (or
the operators of the nonzero spin currents) in the Hamiltonian (which are the
total time derivatives and do not change the classical equations of motion 
but rather affect the topological properties, like the choise of the
$\theta$-vacuum in Haldane's approach) is the reason why the low-lying spin
excitations (and particles for lattice QFT) for our class of models are {\em
gapless} and our low energy theories are conformal. It has to be mentioned
that recent results of the perturbative RG analysis of the zigzag spin
${1\over2}$ chain {\em without} three-spin terms shows the tendency the RG
currents flow to the state with the parity and time-reversal violation
\cite{NGE}. By the way, one can obviously see that the XY limit of the
two-chain spin model does not correspond to the free fermion point of the
exactly solvable model, and this coincides with the results of 
Ref.~\cite{NGE}. Note, though, that in the latter it was erroneously
concluded that the time-reversal and parity symmetries were violated by the
two-chain zigzag spin Hamiltonian {\em with only two-spin couplings} (i.e. the
nearest and next-nearest-neighbor interactions in the single chain picture), 
{\em without spin current terms} in the Hamiltonian. Hence the symmetry of 
the considered state was lower than the symmetry of the Hamiltonian there.

Naturally, the relevant perturbations to our integrable models will
immedeately produce spin gaps. As usual, the algebraic (power-law) decay
of the correlation functions in the groundstate of the models considered in
this paper determines the {\em quantum criticality}. This means that, starting
from the (conformal) exact solutions obtained in this paper one can argue
that the response of the more realistic spin systems to perturbations can be
evaluated by using perturbative methods, e.g., in a renormalization group
framework. For example, let us study the effect of relevant perturbations to
the Hamiltonians considered, ${\hat H}_r = {\hat H} + \delta{\hat H}_1$,
where one can choose as ${\hat H}_r$, e.g., the standard Heisenberg or
uniaxial Hamiltonians for several coupled quantum spin chains, and as
${\hat H}$ the Hamiltonians of spin chains considered exactly in this paper for
some values of $\theta$, where the three-spin terms are relevant. The
correction to the ground state energy and the excitation gap (mass of the
particle in QFT) for the quantum critical system are: $\Delta E \propto -
\delta^{(d+z)/y}$, and $m \propto \delta^{1/y}$, respectively, where $d$ is
the dimension of the system, and $z$ is the dynamical critical exponent. For
the conformally invariant systems studied here one has $d=z=1$. The
application of the standard scaling relations yields $y + x = 2 (=z+d)$,
where $x$ is the scaling dimension, i.e. $x = 2\Delta_l + 2\Delta_r$, found
in the previous sections (for the phases with the dressed charge matrices the
summation over upper indices is meant). Hence the gap for the low-lying
excitations (the mass of the physical particles in QFT) for the perturbed
systems will be $m \propto \delta^{1/2(1-\Delta_l-\Delta_r)}$. Note that 
because of scaling, the behavior of the critical exponents (which are related
to the exponents we introduced for the integrable multi-chain spin models)
in the vicinities of the lines of the phase transitions has to be universal,
and this can be checked experimentally. We expect that the spin gap has to 
exist for the values of the isotropic zigzag inter-chain coupling higher or 
of order of 0.5 for the two-chain spin  ${1\over2}$ system \cite{Zv3}, where 
the three-spin couplings are relevant and the emergence of the spin gap is 
known exactly \cite{MG}.

Very recently, the density matrix renormalization group numerical studies
of the two-chain zigzag spin ${1\over2}$ model (without chiral three-spin
terms in the Hamiltonian) were performed \cite{OHA}. These numerical studies
strongly support the picture proposed here (see also Ref.~\cite{Zv3}):
the magnetization as function of the magnetic field in the groundstate reveals
(i) one second order phase transition (to the spin-saturation phase) for the
weak intra-chain coupling; (ii) one more second order phase transition
between the magnetic (gapless) phases in the intermediate region of the
intra-chain coupling and (iii) in addition to those second order phase
transitions, one to the gapful phase with zero magnetization (plateau) for 
the intra-chain coupling value of 0.5.

We should also mention that it is not the chiral spin terms (as implied in
Ref.~\cite{FR}) but the intra-chain coupling that is responsible for the
commensurate-incommensurate phase transitions between the gapless phases in
this class of models. As for the aforementioned spin currents, their
``fine-tuned'' values produce the cancellation of the spin gap for zero
magnetic field \cite{Zv3}. We should also note that to our mind some
features of the phase diagram obtained in Ref.~\cite{FR1} are artifacts of 
the {\em small number of sites} involved into the numerical
calculations. In Fig.~5 of Ref.~\cite{FR1} in the regions of $0.52 <
\kappa < 0.6$ (corresponding to intra-chain couplings, normalized to the
value of the inter-chain interaction, in the domain [0.54--0.75]) we can
obviously see that when increasing the value of the magnetic field one goes 
from the gapped phase with zero magnetization into the gapless one with two 
Dirac seas of the low-lying excitations, then reaches the gapless phase with 
one Dirac sea, then {\em returnes} to the gapless phase with two Dirac seas, 
and finally reaches the spin-saturated phase. To our mind this return to the
already passed phase is non-physical. One can clearly see that the region
for the values of the intra-chain couplings, where these strange returns
happen, is reduced when going from 16 sites in numerical calculations to 20 
sites. This confirms that presently achieved sizes of the quantum systems for 
numerical calculations can produce even qualitatively invalid results, and 
analytic calculations are necessary, too.

We point out that despite the fact that the relevant perturbations in general
produce a gap for the low-lying excitations, one can apply the results of
this paper to the real gapless multy-chain spin systems, too. For example, it
was recently observed that even for the two-leg ladder system
SrCa$_{12}$Cu$_{24}$O$_{41}$ the spin gap collapses under pressure.
\cite{exp2}

\section{\bf Concluding remarks}

In this paper, motivated by recent progress in the experimental measurements 
for multi-chain spin systems, we have theoretically studied the behavior in 
an external field of a wide class of the multi-chain quantum spin
models and quantum field theories. First, we have investigated the external
field behavior of the exactly integrable two-chain spin ${1\over2}$ model and
have shown that the inclucion of the {\em magnetic anisotropy} of the
``easy-plane'' type, with which the system stays in the quantum critical
region, does not qualitatively change the behavior in an external magnetic 
field. However, we have shown that the magnetic anisotropy {\em changes the 
critical values} of the magnetic fields and intra-chain couplings, at which 
the phase transitions occur, and affects the critical exponents. We have 
shown that the external-field-induced phase transitions we discussed are the 
commensurate-incommensurate phase transitions due to the 
next-nearest-neighbor two-spin interactions, which are present in these
multi-chain models with zigzag-like couplings. We have pointed out that the
low-lying excitations of the conformal limit of our class of multi-chain spin
models are not independent in the incommensurate phase, because they are
governed by the same magnetic field. We have shown that these two-chain
quantum spin models {\em share the most important features} of the behavior
in an external field with the wide class of the (1+1) {\em quantum field
theories}.

We have introduced {\em higher-spin} versions of the two-chain exactly 
solvable spin models, e.g., we have investigated the important class
of 1D two-chain {\em quantum ferrimagnets} with different spin values in the
sites of each chain. Here we have shown that the phase transitions in an
external magnetic field in this exactly solvable two-chain quantum
ferrimagnet are similar in nature to the phase transitions between the
spin-compensated and uncompensated phases in ordinary classical 3D
ferrimagnets.

We have also studied the behavior of the {\em multi-chain} exactly solvable
spin models in an external magnetic field, and shown how the additional phase
transitions arising due to the increasing number of chains vanish in
the quasi-2D limit. Hence, to the best of our knowledge, we have proposed the
first exact scenario of the transition from 1D to 2D quantum spin models in 
the presence of an external magnetic field. We have argued that the
commensurate-incommensurate phase transitions in the multi-chain quantum
spin models have to disappear in the limit of an infinite number of chains.

Finally, we have shown how the relevant deviations from the integrability,
e.g., the absence of the three-spin (spin chiral) terms in the Hamiltonians,
which separately break the parity and time-reversal symmetries, give rise to 
{\em gaps} in spectra of the low-lying excitations of the multi-chain quantum 
spin systems and we have calculated the critical scaling exponents for these 
gaps. We pointed out the qualitative agreement of our exact analytic 
calculations with recent numerical simulations for zigzag spin models.

I am grateful to A.~G.~Izergin, S.~V.~Ketov , A.~Kl\"umper, V.~E.~Korepin,
G.~I.~Japaridze, A.~Luther  and A.~A.~Nersesyan for helpful discussions. I
thank J.~Gruneberg for his kind help. The financial support of the Deutsche
Forschungsgemeinschaft and Swedish Institute is acknowledged.


\begin{thebibliography}{99}
\bibitem{ER} For a recent review on so-called ``ladder'' systems see,
E.~Dagotto and T.~M.~Rice, Science {\bf 271}, 618 (1996) and references
therein.
\bibitem{exp} D.~C.~Johnston, J.~W.Johnston, D.~P.~Goshorm and A.~P.Jacobson,
Phys. Rev. B {\bf 35}, 219 (1987); Z.~Hiroi, M.~Azuma, M.~Takano and
Y.~Bando, J. Sol. St. Chem. {\bf 95}, 230 (1990); M.~Azuma, Z.~Hiroi,
M.~Takano, K.~Ishida and Y.~Kitaoka, Phys. Rev. Lett. {\bf 73}, 3463 (1994);
Y.~Ajiro, T.~Asano, T.~Inami, H.~Aruga-Katori and T.~Goto, J. Phys. Soc.
Jpn. {\bf 63}, 859 (1994); G.~Chaboussant P.~A.~Crowell, L.~P.~L\'evy,
O.~Piovesana, A.~Madouri and D.~Mailly, Phys. Rev. B {\bf 55}, 3046 (1997);
S.~A.~Carter, B.~Batlogg, R.~J.~Cava, J.~J.~Krajewski, W.~F.~Peck,~Jr. and
T.~M.~Rice, Phys. Rev. Lett. {\bf 77}, 1378 (1996); G.~Chambourssant,
Y.~Fagot-Revurat, M.-H.~Julien, M.~E.~Hanson, C.~Berthier, M.~Horvati\'c,
L.~P.~L\'evy and O.~Piovesana, Phys. Rev. Lett. {\bf 80}, 2713 (1998);
W.~Shiramura, K.~Takatsu, B.~Kurniawan, H.~Tanaka, H.~Uekusa, Y.~Ohashi,
K.~Takizawa, H.~Mitamura and T.~Goto, J. Phys. Soc. Jpn. {\bf 67}, 1548
(1998).
\bibitem{Tsv} A.~M.~Tsvelik, Phys. Rev. B {\bf 42}, 779 (1990).
\bibitem{PopZv} V.~Yu.~Popkov and A.~A.~Zvyagin, Phys. Lett. {\bf 175A}, 295
(1993).
\bibitem{MT} N.~Muramoto and M.~Takahashi, preprint, cond-mat/9902007.
\bibitem{Bet} H.~Bethe, Z. Phys. {\bf 71}, 205 (1931).
\bibitem{qism} V.~E.~Korepin, N.~M.~Bogoliubov, and A.~G.~Izergin, {\em
Quantum Inverse Scattering Method and Correlation Functions}, Cambridge
University Press, Cambridge, 1993.
\bibitem{OYA}M.~Oshikawa, M.~Yamanaka and I.~Affleck, Phys. Rev. Lett. 
{\bf 78}, 1984 (1997).
\bibitem{Zv3} A.~A.~Zvyagin, Phys. Rev. B {\bf 57} 1035 (1998).
\bibitem{FR} H.~Frahm and C.~R\"odenbeck, J. Phys. A {\bf 30}, 4467 (1997).
\bibitem{Zv1} A.~A.~Zvyagin, Pis'ma v Zh. Eks. Teor. Fiz. {\bf 60}, 563
(1994); [JETP Lett. {\bf 60}, 580 (1994)]; Phys. Rev. B {\bf 51}, 12579
(1995); A.~A.~Zvyagin and H.~Johannesson, Europhys. Lett. {\bf 35}, 151
(1997).
\bibitem{Bax} R.~J.~Baxter {\em Exaxtly Solved Models in Statistical
Mechanics}, Academic, Orlando, 1982.
\bibitem{Zv2} A.~A.~Zvyagin, Pis'ma v Zh. Eksp. Teor. Fiz. {\bf 63}, 192
(1996) [JETP Lett. {\bf 63}, 204 (1996)].
\bibitem{AL1} N.~Andrei and J.~H.~Lowenstein, Phys. Rev. Lett. {\bf 43},
1698 (1979).
\bibitem{FT} L.~D.~Faddeev and L.~A.~Takhtadjan, Phys. Lett. {\bf 85A}, 375
(1981).
\bibitem{FKM} K.~Fabricius, A.~Kl\"umper and B.~M.~McCoy, preprint
cond-mat/9810278.
\bibitem{cft} A.~A.~Belavin, A.~M.~Polyakov and A.~B.~Zamolodchikov, Nucl.
Phys. {\bf B241}, 333 (1984). See, also, J.~L.~Cardy, Nucl. Phys. {\bf B270},
186 (1986); H.~W.~J.~Bl\"ote, J.~L.~Cardy and M.~P.~Nightingale, Phys. Rev.
Lett. {\bf 56}, 742 (1986); I.~Affleck, {\em ibid.} {\bf 56}, 746 (1986).
\bibitem{conf} H.~J.~de~Vega and F.~Woynarovich, Nucl. Phys., {\bf B251}, 439
(1985); N.~M.~Bogoliubov, A.~G.~Izergin and V.~E.~Korepin, Nucl. Phys. {\bf
B275}, 687 (1986); F.~Woynarovich and H.-P.~Eckle, J. Phys. A {\bf 20}, L97
(1987); {\em ibid.} {\bf 20}, L443 (1987); N.~M.~Bogoliubov, A.~G.~Izergin and
N.~Yu.~Reshetikhin, J. Phys. A {\bf 20}, 5361 (1987); A.~G.~Izergin,
V.~E.~Korepin and N.~Yu.~Reshetikhin, J. Phys. A {\bf 22}, 2615 (1989);
F.~Woynarovich, H.-P.~Eckle and T.~T.~Truong, J. Phys. A {\bf 22}, 4027
(1989); H.~Frahm and V.~E.~Korepin, Phys. Rev. B {\bf 42}, 10553 (1990);
N.~Kawakami and S.-K.~Yang, Phys. Rev. Lett. {\bf 65}, 2309 (1990);
H.-P.~Eckle and C.~J.~Hamer, J. Phys. A {\bf 24}, 191 (1991).
\bibitem{FR1} H.~Frahm and C.~R\"odenbeck, preprint, cond-mat/9812103.
\bibitem{Kor} V.~E.~Korepin, Theor. Math. Phys. {\bf 41}, 953 (1979).
\bibitem{PL} S.~Park and K.~Lee, J. Phys. A {\bf 31}, 6569 (1998).
\bibitem{JN} G.~I.~Japaridze and A.~A.~Nersesyan, Phys. Lett. {\bf 85A}, 23
(1981).
\bibitem{DS} C.~Destri and T.~Segalini, Nucl. Phys. {\bf B 455}, 759 (1995).
\bibitem{DdV2} C.~Destri and H.~J.~de~Vega, Phys. Lett. {\bf 201B}, 245
(1988).
\bibitem{LE} A.~Luther and V.~J.~Emery, Phys. Rev. Lett. {\bf 33}, 589 (1974).
\bibitem{TTF} V.~O.~Tarasov, L.~A.~Takhtadzhyan and L.~D.~Faddeev, Theor.
Math. Phys. {\bf 57}, 1059 (1983).
\bibitem{BI} N.~M.~Bogolybov and A.~G.~Izergin, Theor. Math. Phys. {\bf 59},
441 (1984).
\bibitem{DdV1} C.~Destri and H.~J.~de~Vega, J. Phys. A {\bf 22}, 1329 (1989).
\bibitem{JNW} G.~I.~Japaridze, A.~A.~Nersesyan, and P.~B.~Wiegmann, Nucl.
Phys. {\bf B230}, 511 (1984).
\bibitem{DdV3} C.~Destri and H.~J.~de~Vega, Nucl. Phys. {\bf B504}, 621 (1997).
\bibitem{AL2} N.~Andrei and J.~H.~Lowenstein, Phys. Lett. {\bf 90B}, 106
(1980).
\bibitem{PW1} A.~M.~Polyakov and P.~B.~Wiegmann, Phys. Lett. {\bf 131B}, 121
(1983).
\bibitem{PW2} A.~M.~Polyakov, and P.~B.~Wiegmann, Phys. Lett. {\bf 141B}, 223
(1984).
\bibitem{Wie2} P.~B.~Wiegmann, Phys. Lett. {\bf 141B}, 217 (1984);
{\em ibid}, {\bf 152B}, 209 (1985).
\bibitem{TaBa} L.~A.~Takhtajan, Phys. Lett. {\bf A87}, 479 (1982);
H.~M.~Babujian, Nucl. Phys. B {\bf 215} 317 (1983).
\bibitem{dVW} H.~J.~de Vega and F.~Woynarovich, J. Phys. A {\bf 25}, 4499
(1992).
\bibitem{Sch} P.~Schlottmann, Phys. Rev. B {\bf 49}, 9202 (1994).
\bibitem{ZS} A.~A.~Zvyagin and P.~Schlottmann, Phys. Rev. B {\bf 52}, 6569
(1995).
\bibitem{KRS} P.~P.~Kulish, N.~Yu.~Reshetikhin, and E.~K.~Sklyanin, Lett.
Math. Phys. {\bf 5}, 393 (1981).
\bibitem{KB} L.~Kadanoff and A.~C.~Brown, Ann. Phys. NY {\bf 121}, 318 (1979).
\bibitem{par1} A.~B.~Zamolodchikov and V.~A.~Fateev, Sov. Phys. JETP {\bf
62}, 215 (1985); D.~Gepner and Z.~Qiu, Nucl. Phys., {\bf B285}, 423 (1987).
\bibitem{par2} A.~N.~Kirillov and N.~Yu.~Reshetikhin, J. Phys. A {\bf 20},
1587 (1987); F.~C.~Alcaraz and M.~J.~Martins, J. Phys. A {\bf 22}, 1829
(1989); H.~Frahm and N.-C.~Yu, J. Phys. A {\bf 23}, 2115 (1990).
\bibitem{SchZv} P.~Schlottmann and A.~A.~Zvyagin, Phys. Rev. B {\bf 56} 13989
(1997).
\bibitem{LSM} E.~Lieb, T.~Schultz and D.~Mattis, Ann. Phys. (NY) {\bf 16},
407 (1961). See also I.~Affleck and E.~H.~Lieb, Lett. Math. Phys. {\bf 12}, 
57 (1986).
\bibitem{Hal} F.~D.~M.~Haldane, Phys. Lett. {\bf A93}, 464 (1983).
\bibitem{NGE} A.~A.~Nersesyan, A.~O.~Gogolin and F.~H.~S.~Essler, Phys. Rev.
Lett. {\bf 81}, 910 (1998).
\bibitem{MG} C.~K.~Majumdar and D.~K.~Ghosh, J. Math. Phys. {\bf 10}, 1388
(1969).
\bibitem{OHA} K.~Okunishi, Y.~Hieida and Y.~Akutsu, preprint, cond-mat/9904155.
\bibitem{exp2} H.~Mayaffre, P.~Auban-Senzier, M.~Nardone, D.~J\'erome,
D.~Poilblanc, C.~Bourbonnais, U.~Ammerahl, G.~Dhalenne and A.~Revcolevschi,
Science {\bf 279}, 345 (1998).
\end{thebibliography}
\end{document}